\documentstyle[12pt,epsfig]{article}
\begin{document}


\begin{center}
{\large \bf Limits on the magnetic moment of sterile neutrino and two-photon 
neutrino decay} 
\end{center}
\vspace{0.5cm}

\begin{center}  
S.N.~Gninenko\footnote{E-mail address:
 Sergei.Gninenko\char 64 cern.ch} and 
N.V.~Krasnikov\footnote{nkrasnik\char 64 vxcern.cern.ch}\\
{\it Institute for Nuclear Research of the Russian Academy of Sciences,\\ 
Moscow 117312}
\end{center}

\begin{abstract} It is shown that the  non-zero transition magnetic moment 
($\mu_{tran}$) between the 
sterile neutrino ($\nu_{s}$) and the muon neutrino ($\nu_{\mu}$) could be effectively searched for via the Primakoff effect, 
in the process of $\nu_{\mu} Z \rightarrow \nu_{s}Z$
 conversion in the external Coulomb field of a nucleus $Z$, with the subsequent 
 $\nu_{s}\rightarrow \nu_{\mu} + \gamma$ decay.\  From
the recent results of the NOMAD neutrino detector at CERN  
a model-independent constraint of 
 $\mu_{tran} < (10^{-6} - 10^{-9}) \mu_{B}$ is obtained depending on the value
of $\nu_{s}$ mass.\ For the $m_{s}\sim O(1)~ GeV$ region these bounds are comparable with
 the present experimental ones on $\nu_{\mu}$ and 
$\nu_{e}$ diagonal magnetic moments and are more sensitive than those on 
$\nu_{\tau}$ magnetic moment.\    

From the same analysis the constraint on 
$\nu_{\tau}(\nu_{s})\rightarrow \nu_{\mu} +\gamma + \gamma$ decay lifetime 
$\tau > 2\times10^{13} sec/m_{\nu}^{7}(MeV)$  
  is obtained.\
The limit is valid for neutrino masses
up to $m_{\nu}\sim O(1)~GeV$.\
\end{abstract}

\section{Introduction}
  Over the past few years it has been  realized that a new fourth neutrino 
in addition to the three known neutrinos is highly required in  many 
extensions of the Standard Model, e.g. see \cite{1}.\
Neutrinos of new flavour with a mass smaller than $M_{Z}/2$ cannot be a 
fourth $SU(2)_{L}$ doublet 
of left-handed neutrino, since they are excluded by invisible width of 
$Z$ boson \cite{2}.\
They have to be SU(2) singlet (sterile) neutrinos,
 which do not interact through $W^{\pm}$ or 
$Z$ bosons of the Standard Model.\

 The light  
sterile neutrino ($m_{\nu_{s}} \leq 1~ eV$) seems unavoidable to explain   
significant anomalies which have been observed in experiments on solar neutrinos
\cite{3} -\cite{7}, atmospheric neutrinos \cite{8} -\cite{10} and accelerator neutrinos 
\cite{11}.\
The heavy neutrino  ($m_{\nu_{s}} \geq 1~ TeV$) could help to solve the problem of the origin of the 
excess of baryons over anti-baryons in the Universe.\ 
The baryon asymmetry could originate from the lepton asymmetry due to 
non-pertubative electroweak effects, which were generated in the early Universe
via CP- and lepton number non-conserving decays of heavy Majorana neutrinos \cite{12}.\
   In the recent publication \cite{13} a new mechanism of leptogenesis 
was proposed.\ It suggests that asymmetries in lepton numbers were generated 
due to oscillations of heavy ( $m_{\nu_{s}} = 1 - 100~ GeV$) 
sterile neutrinos and their interactions with 
ordinary matter in the early Universe.\
Such neutrinos could mix with the 
standard neutrinos or decay into them.\
Another interesting example is related to the explanation of the KARMEN 
anomaly \cite{14} as the decay signature of the 33.9 MeV/c$^2$ sterile
 neutrino \cite{15}.\
One of the possible mechanisms of the sterile neutrino  decays 
for the case of KARMEN anomaly could be due to a  
 non-zero transition magnetic moment of the sterile neutrinos.\ Thus,  
 the search for sterile neutrinos in the MeV - GeV 
range through their electromagnetic interactions is 
rather important and interesting.\\  

Non-zero electromagnetic properties
of the neutrino have been discussed in many extensions of the Standard Model 
, for review see e.g. \cite{16}.\ Usually large magnetic moment imply large 
masses.\ 
In the simplest extension, for example, 
neutrino masses and corresponding  magnetic moments (acquiring  through 
radiative corrections) are proportional \cite{17} :
\begin{equation}
2\mu_{ij} = \frac{3eG_{F}m_{ij}}{8\sqrt{2} \pi^{2}} = 3.2\times 10^{-19
}( m_{ij}/1~eV) \mu_{B} 
\end{equation}  
where $G_{F}$ is the Fermi constant, $\mu_{ij} (i,j = e, \mu, \tau )$ is 
the neutrino magnetic moment matrix,
 $m_{ij}$ is the Dirac neutrino mass matrix and $\mu_{B} = e/2m_{e}$ is
 the Bohr magneton.\  
  
The interesting electromagnetic processes  with the participation
of sterile neutrinos, if they exist, and photons  
are single or  two photon neutrino decays $\nu_{i} \rightarrow \nu_{j} + 
\gamma $, $\nu_{i} \rightarrow \nu_{j} + 
\gamma + \gamma $ or  neutrino-photon 
scattering $\nu_{i} + \gamma \rightarrow \nu_{j} + \gamma $, where 
$\nu_{i}$ stands also for $\nu_{s}$.\ Note that in 
general one and two photon neutrino decays are independent.\ For instance, 
in the Standard Model  with massive neutrino for neutrino mass 
$m_{\nu_{i}} \sim O(1)~ MeV$ the two photon decay 
is not GIM suppressed and can dominate \cite{18}.\
Therefore, it would be interesting 
to obtain direct experimental bounds on the two 
photon neutrino decay and on the neutrino photon scattering.\

The most stringent experimental bounds on magnetic moments of $\nu_{\mu}$
and $\nu_{e}$ neutrinos, 
$\mu_{\nu_{e}} < 1.1\times 10^{-10}\mu_{B}$ and $\mu_{\nu_{\mu}} < 7.4\times 10^{-9}\mu_{B}$, were obtained in ref. \cite{19} from the analysis of 
$\nu-e$ scattering.\ The limits on magnetic moments of $\nu_{\tau}$ are 
less stringent: $\mu_{\tau\tau} < 5.4\times 10^{-7}\mu_{B}$ 
 \cite{20} and 
$\mu_{\tau x} < 10^{-9}\times(MeV/m_{\nu_{\tau}})^{2}\mu_{B}$ \cite{21}
from measurements at BEBC,
$\mu_{\tau x} < 4.0\times 10^{-6}\mu_{B}$ from the  
LEP data \cite{22} and 
$\mu_{\tau\tau}$(or $\mu_{\tau x}$) $< 3.3\times 10^{-6}\mu_{B}$
from the recent analysis of single photon production by the L3 Collaboration
\cite{23}.\
The magnetic moment of the sterile neutrino could be 
constrained from the supernova SN1987a observations.\ However, the bounds are 
model-dependent \cite{24} and will not be considered further here.\

In this Letter we apply the Primakoff idea to measure the
$\pi^0\rightarrow2\gamma$ width by measuring the cross section
for the scattering process $\gamma Z\rightarrow \pi^{0} Z$ 
for measurements of radiative neutrino decays.\ Here $Z$ stands for a heavy 
nucleus.\
We show that  non-zero transition magnetic moment $\mu_{tran}$ between 
  sterile neutrino $\nu_{s}$ and muon neutrino could be effectively 
searched for via the Primakoff effect, 
in the process of $\nu_{\mu} Z \rightarrow \nu_{s}Z$
 conversion in the external Coulomb field of a nucleus with the subsequent 
decay of $\nu_{s}\rightarrow \nu_{\mu} + \gamma$.\ Hereinafter, it is assumed 
that mixing is small and the flavour eigenstates are defined by their 
primary mass eigenstates:  $\nu_1\simeq\nu_e$, $\nu_2\simeq\nu_{\mu}$,
$\nu_3\simeq\nu_{\tau}$ and $\nu_4\simeq\nu_s$.\ So, that the transition
$\nu_{2} Z \rightarrow \nu_{4}Z$ would appear at the experimental level as
$\nu_{\mu} Z \rightarrow \nu_{s}Z$.\   
An estimate of the bounds on  transition magnetic moments $\mu_{tran}$
 is obtained
from the recent results of the NOMAD neutrino detector at CERN \cite{25}.\  
Note that these bounds would also apply to any other $\nu_i$ which mixes 
significantly in $\nu_{\mu}$ or $\nu_s$.
From the same analysis,
we  deduce also a bound on the width of 
two photon $\nu_{\tau}(\nu_{s})\rightarrow \nu_{\mu} \gamma \gamma $  neutrino decay 
via the use of Primakoff reaction $\nu_{\mu} Z \rightarrow \nu_{\tau}(\nu_{s})
+ \gamma +Z$.\ In both cases a photon from neutrino decay is played by the 
Coulomb field of the nucleus (see for example Fig.1).\ 

It should be noted that the use of the magnetic 
Primakoff effect has been discussed in refs. \cite{26,27} for the 
investigation of the one-photon neutrino decay 
$ \nu_{i} \rightarrow \nu_{j} + \gamma $.\ The present work is based mainly 
on result obtained in ref. \cite{28}.\
  
The organisation of the paper is the following. In section 2 we 
give the formulae for the Primakoff
cross sections and the decay widths for the neutrinos with non-zero transition 
magnetic moments. In section 3 we give the formulae for the decay widths 
and the cross sections for the processes  $\nu_{i} \rightarrow \nu_{j} +
\gamma \gamma$ and $\nu_{j}Z \rightarrow \nu_{i}\gamma  Z$.  
In section 4 we estimate the 
bounds on the transition magnetic moments and  on two photon neutrino decay 
widths from the recent NOMAD results.\ Section 5 contains concluding 
remarks.  

\begin{figure}[hbt]
\begin{center}
   \mbox{\hspace{-1.5cm}\epsfig{file=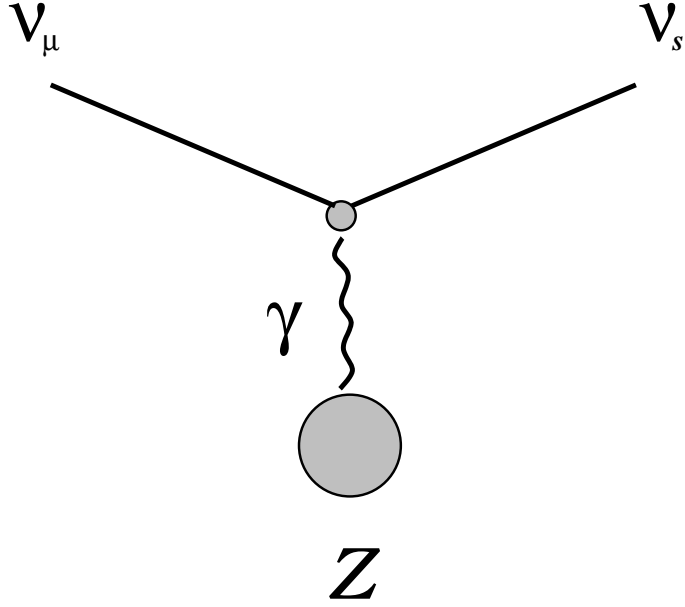,height=80mm}}
  \caption{\em Production of the sterile  neutrino by the
Primakoff effect with subsequent radiative neutrino decay 
$ \nu_{s} \rightarrow \nu_{\mu} + \gamma $.}     
\end{center}
  \label{figure 19:}
\end{figure}
 
\section{Primakoff cross section for $\nu_{j} Z \rightarrow \nu_{i} Z$}

The Lagrangian describing the neutrino interaction with electromagnetic 
field due to non-zero anomalous transition magnetic moment has the form 

\begin{equation}
L_{\mu} = \mu_{tran} \bar{\nu}_j \sigma_{\mu \nu} \nu_{i} F_{\mu \nu}/2 +h.c.
\end{equation}

Here for simplicity we consider the case when $\nu_{i}$ and $\nu_{j}$ 
are Dirac neutrinos with masses $m_i$ and $m_j$, respectively. The corresponding modification for the case 
of Majorana neutrinos is straightforward. The decay width $\nu_{i} 
\rightarrow \nu_{j} + \gamma$ is determined by the formula{\footnote {For the 
case of nonzero transition electric moment $d_{tran}$we have to replace in Eq.(3)
(also in Table 1) $|\mu_{tran}|\rightarrow |\mu_{tran} - d_{tran}|$}} 
\begin{equation}
\Gamma(\nu_{i} \rightarrow \nu_{j} + \gamma) = 
\frac{m_{i}^{3} \mu^{2}_{tran}}  {8\pi}
(1 -\frac{m_j^2}{m_i^2})^3~~,
\end{equation}
where $\mu_{tran}$ is the transition magnetic moment between $i,j$ neutrino states.\
The cross section for the reaction $\nu_{j} Z \rightarrow \nu_{i} Z$ 
through the Primakoff effect  has the form
\begin{equation}
\frac{d\sigma}{d \Omega}(\nu_j Z \rightarrow \nu_i Z) = 
\frac{8\Gamma(\nu_i \rightarrow \nu_j \gamma)}{m^2_i \Delta} 
\alpha Z^2 P^4_{LAB} 2(1 -\cos{\theta})\frac{F^2(t)}{t^2},
\end{equation}
\begin{equation}
\sigma = \frac{8\pi \Gamma(\nu_i \rightarrow \nu_j \gamma)}{m^3_i\Delta} 
\alpha Z^2 \int^{\infty}_{t_{min}}\frac{(t-t_{min})F^2(t)}{t^2}dt .
\end{equation}
Here $\Delta = (1 - \frac{m^2_j}{m^2_i})^2$, $t = - (p_{\nu_{i}} - 
p_{\nu_{j}})^2$, $t_{min} = (\frac{m^2_i - m^2_j}{2P_{LAB}})^2$, 
$P_{LAB}$ is the momentum of the incoming neutrino in the laboratory 
system and $F^{2}(t)$ is form-factor of the target.\ In our  estimates we shall use the model of form-factors 
described in ref. \cite{29}.

\section{Cross section and decay width for two-photon processes}

Let us now consider the two-photon neutrino decay 
$\nu_{i}\rightarrow \nu_{j} +\gamma + \gamma$.\
One can find that the Primakoff cross section for the reaction $\nu_{j} 
+ Z \rightarrow 
\nu_{i} + \gamma + Z$ in the approximation of the equivalent photons 
for the scattering of neutrino on heavy nuclei is given 
in the laboratory frame by the formulae:
 
\begin{equation}
\sigma(\nu_{j} Z  \rightarrow \nu_{i} \gamma Z) = 
\frac{\alpha Z^2}{4P^{2}_{LAB}} \int \Phi_{1}(t) 
\frac{F^2(t)}{t^2}\frac{t-t_0}{t} dt,
\end{equation}
\begin{equation}
\Phi_{1}(t) = \int _{0}^{2P_{LAB} \sqrt{t}} \Phi(x) dx ,
\end{equation}
where $Z^2 F^2(t)$ is the target form-factor, t is 
the square of the momentum 
transfer, $t_0$ is the minimal transfer momentum square. 
$ \Phi(x) = x \sigma(x)$ and $\sigma(s)$ is the cross section 
for the reaction $\nu_{i}\gamma \rightarrow \nu_{j} \gamma$. Unfortunately 
Eq.(6) is not very useful for deriving  bounds on 
neutrino-photon interaction.\ To obtain more useful  formulae we 
parametrise the neutrino-photon scattering using the method of the 
effective Lagrangians. The two-photon neutrino decay width or photon-neutrino 
scattering is described by d=7 effective Lagrangian: 

 \begin{equation}
L = \sum_{k=1}^2 [\bar{\psi}_{j\nu}(F_{Sk,ij} + F_{Ak,ij}
\gamma_{5}]\psi_{i\nu} + 
h.c.]A_{k} ,
\end{equation}
\begin{equation}
A_1 =\frac{1}{4}\epsilon^{\mu \nu \alpha \beta}F_{\mu \nu}
F_{\alpha \beta},
\end{equation}
\begin{equation}
A_2 = \frac{1}{2} F_{\mu\nu}F^{\mu\nu},
\end{equation}
where the coefficients $F_{Sk,ij}$, $F_{Ak,ij}$ have the dimension $[mass]^{-3}$. 
Here we again consider the case of the Dirac neutrino.\ The generalisation to 
the case of the Majorana neutrino is straightforward.\
The Lagrangian (8) is the most general d=7 Lagrangian describing two-photon
 neutrino decay.\ The d=8 operators are suppressed by additional 
power of high inverse mass and we shall ignore them.  For the effective 
Lagrangian (8) the cross-section for the reaction 
$\nu_{j}Z \rightarrow \nu_{i} + \gamma +Z$ has the form
\begin{equation}
\frac{d\sigma(\nu_{j}Z \rightarrow \nu_{i}\gamma Z)}{dtdk_3} = 
\frac{\alpha |\vec{k}|^3}{16\pi^{2}P^2_{LAB}}Z^2F^2(t) ,
\end{equation}
\begin{equation} 
\sigma(\nu_{j}Z \rightarrow \nu_{i} \gamma Z) = 
\int \frac{\alpha Z^2P_{LAB}^2}{64\pi^{2}}K_{ij}F^2(t)dt,
\end{equation}
\begin{equation}
K_{ij} = \sum^{2}_{k=1}[|F_{Sk,ij}|^2 + |F_{Ak,ij}|^2] \equiv 
\frac{1}{M^6_{eff}} .
\end{equation}

For the effective Lagrangian (8) the differential decay width rate 
for the reaction $ \nu_{i} \rightarrow \nu_{j} \gamma \gamma $ is 
\begin{equation}
\frac{d\Gamma}{ds} = \frac{1}{256\pi^{2}}
\frac{s^2F(s) \lambda^{\frac{1}{2}}
(s,m^2_i,m^2_j)}{m^3_i},
\end{equation}
\begin{equation}
\lambda(x,y,z)=(x-y-z)^2 -4yz,
\end{equation}
\begin{equation}
F(s) = (|F_{Si,ij}|^2 +|F_{Sj,ij}|^2)((m_i+m_j)^2-s) +
(|F_{Ai,ij}|^2 + |F_{Aj,ij}|^2)((m_i-m_j)^2-s)
\end{equation}
Here $s =(k_i+k_j)^2$ is the invariant mass square of the photon pair 
($0 \leq s \leq (m_i -m_j)^2$.
For the most interesting case  $m_i \gg m_j$ we have 
\begin{equation}
\Gamma(\nu_{i} \rightarrow \nu_{j} \gamma \gamma) = 
\frac{m^7_i(MeV)}{M^6_{eff}(GeV)}\times 5\times10^{-3} sec^{-1}
\end{equation}

\section{Method of Search at NOMAD}

In this section we consider the NOMAD neutrino detector \cite{30} 
as an example in order to estimate bounds on $\nu_{s}$ transition magnetic 
moments and on two photon neutrino decay width.\ The NOMAD detector, 
designed to search for a neutrino oscillation signal in the CERN SPS 
wide-band neutrino beam, is described in detail in ref. \cite{30}.\ 
The neutrino beam with the average muon neutrino momentum 
$<P_{\nu_{\mu}}> = 27~ GeV$ is generated by 450 GeV protons delivered by 
the SPS to the Be neutrino target. 
Consider the production of sterile neutrino in  NOMAD
(we will follow bellow  to similar estimate of the NOMAD sensitivity 
 for a new light boson search described in ref. \cite{28}).\

The experimental signature for search for  the single photon (see Fig.1) or
two-photon radiative neutrino decay is the single high energy gamma quantum 
 with the average energy 
$E_{\gamma} \approx \frac{E_{\nu_{\mu}}}{2}$ in the forward direction 
that results in a single isolated electromagnetic 
shower in the detector.\ Because the cross sections for 
$\nu_{\mu}Z \rightarrow \nu_{s}Z$  and 
$\nu_{\mu}Z \rightarrow \nu_{s} \gamma Z$ 
are proportional to $Z^2$, preferable search for such events is  in the lead 
of the NOMAD preshower(PRS) detector\cite{30}.\ 

  The occurrence of $\nu_{\mu}Z \rightarrow \nu_{s} \rightarrow 
\nu_{\mu} \gamma Z$ or $\nu_{\mu}Z \rightarrow \nu_{s} + \gamma +Z$ events  would appear as an excess of
neutrino-like interactions in the PRS with pure electromagnetic final 
states above those expected from  Monte Carlo predictions.\ The main contribution to the background is expected from the standard neutrino processes which 
have a significant electromagnetic component in the final state, e.g.
 coherent and diffractive $\pi^{0}$ production; $\nu_{e}CC$ interactions, 
quasi-elastic $\nu_{e}$ scattering, etc..

The results of refs. \cite{25,28} on the NOMAD search for 
the hypothetical X-boson, where the similar signature with the single 
high energy photon in the final state has been used, constrain the  
cross section $\sigma  \equiv \sigma(\nu_{\mu} + Z \rightarrow \gamma + Z + 
+ invisible)$ to 
\begin{equation}
\sigma \leq 10^{-3} pb
\end{equation}

To calculate the Primakoff cross section  on lead ( Eq.(5)), the model of atomic and 
nuclei form-factors described 
in ref. \cite{28} was used.\ The target form-factor $Z^2F^2(t)$ consists 
of three parts. At small $t \leq t_0$ we use the Thomas-Fermi-Moliere model 
for atomic form-factors \cite{29}: 
\begin{equation}
Z^2F^2(t) = G^{el}(t) + G^{inel}(t),
\end{equation}
\begin{equation}
G^{el}(t) = Z^2\frac{a^4t^2}{(1 + a^2t^2)^2} ,
\end{equation}
\begin{equation}
G^{inel}(t) = Z\frac{a^4_1t^2}{(1 +a^2_1t)^2},
\end{equation}
where $ a= 111.7 Z^{-\frac{1}{3}}/m_e$, 
$a_1 = 724.2 Z^{-\frac{2}{3}}/m_e$.\ For values $t \geq t_{0} =7.39m^2_{e}$ we use the elastic nuclear 
form-factor \cite{28}: 
\begin{equation}
Z^2F^2(t) = G^{nucl}(t),\hspace{0.5cm}
G^{nucl}(t) = \frac{Z^2}{(1 + \frac{t}{d})^2} ,
\end{equation}
where $d =0.164A^{-\frac{2}{3}}GeV^2$ and $A$ is the mass number.

These calculations combined with limit of Eq.(18) result in 
bounds to the neutrino transition 
magnetic moment $\mu_{tran}$ which are shown in Table 1 for different 
$\nu_{s}$ masses. 
 In this estimate, based on MC simulations \cite{28}, 
length for neutrino decay was taken to be 
$\sim 1$ cm (thickness of the PRS lead, \cite {30}).\
In the second column the upper bound on $\mu_{tran}/\mu_{B}$ is shown 
for NOMAD detector with $l_0 \sim 1$ cm.  The corresponding exclusion region
for this case is illustrated also in Fig.2.\
Since the length of neutrino decay  considered is $l_{0} \approx 1$ cm 
 only $\sim l_0/c\tau$ fraction of neutrino radiative decays 
is detected. Here, the life time of the ultrarelativistic neutrino is 
determined by the formula $\tau = E_{\nu_{s}}/m_{\nu_{s}}\times 1/\Gamma(\nu_{s}\rightarrow \nu_{\mu} + \gamma)$, $E_{\nu_{s}} 
\approx E_{\nu_{\mu}}$.\ Therefore, the bound of Eq.(18) turns into the
 following bound on the product of the cross section and decay width of the 
sterile neutrino:

\begin{equation}
\sigma(\nu_{\mu} + Z \rightarrow \nu_{s} + Z)\times l_0/c\tau\leq 10^{-3} pb
\end{equation}

 Consider for illusration numerical example for neutrino mass   
 $m_{\nu_{s}} = 1$ MeV ( first raw in Table 1).\ 
For this case  taking into account Eq.(23)  
results in limit $\mu_{tran} < 1.6 \cdot 10^{-6}
\mu_{B}$.\ Indeed,  for $\mu_{tran} = 1.6 \cdot 10^{-6} \mu_{B}$
 the sterile neutrino decay length $c\tau \approx 6 \cdot 10^{7}$ 
cm, the suppression factor $l_0/c\tau \approx 1.7 \cdot 
10^{-8}$, the cross section is 
$\sigma \sim \mu^{2}_{tran} \sim 6.7 \cdot 10^{4}$ pb and the 
product of the cross section times suppression factor 
is $\sigma\times l_0/c\tau \approx 10^{-3}$ pb, which is in agreement with Eq.(23).

Note that the radiative decay of $\nu_{i} \rightarrow  \nu_{j} + \gamma $ 
of a neutrino in 
the Standard Model with lepton mixing is enhanced in the Coulomb field of a 
nucleus \cite{31,32}. For the case of sterile 
neutrino standard GIM suppression 
factor $(m_l/m_W)^4$ is absent \cite{16} and the formula (6) of 
ref.\cite{31} for the decay width enhancement factor has the form
$ R = \frac{\omega}{\omega_{0}} \sim 14 (\frac{E_{\nu_{s}}}{m_{\nu_{s}}})^2 
(\frac{eE}{m^2_{\mu_{s}}})^2 \sin^2(\varphi)$ .
Here, $\omega$ is the neutrino decay probability in the Coulomb field of a 
nucleus, $\omega_{0}$ is the neutrino decay probability in the vacuum, $E$ is 
the electric field of the nucleus and $\varphi$  is an angle between the vector of 
momentum $\vec{p}$ of the decaying neutrino and the electric field strength 
$\vec{E}$. The typical average Coulomb nucleus field is 
$E \sim eZ/4\pi r^2_0$, where $r^2_0 = d^{-1}$.\  Numerically we find that for Pb the enhancement factor is 
$\omega/\omega_{0} \sim 0.1 (1~ GeV/m_{\nu_{s}})^6$ . 
For $m_{\nu_{s}} = 10$ MeV, for example, we find that $\omega/\omega_{0} \sim 
10^{11}$ and the neutrino decay length in the Coulomb nucleus field is $l 
\sim 10^{-3}$ cm.\ However, neutrino spends effectively only small fraction
 of its life time 
$r_{0}/l \sim 10^{-9}$ in the region of the Coulomb field  of the Pb nucleus.\
 This number has to be compared to the 
 fraction of $\sim 10^{-6}$ (see Eq.(3) and Table 1) of neutrino decays in the PRS detector with decay length 
$l_0 \approx 1$ cm.\ So, the enhancement effect is less than $1\%$, for higher masses it is even smaller.\ 
This effect is model dependent, it is maximal for electron propagator in the loop, \cite{31},  anyway 
enhancement effects in neutrino radiative decays only improve our bounds 
(table 1, column 3). \footnote{We are indebted to N.V.Mikheev and 
L.A.Vassilevskaya for explanation of the meaning of refs. \cite{31,32}.}

\begin{table}[h]
\begin{center}
\begin{tabular}{|l||l|}
\hline
$m_{\nu_{s}}$, GeV & $\mu_{tran}/\mu_{B}$ \\
\hline 
0.001   & $1.6 \cdot  10^{-6}$     \\
\hline 
0.01    & $1.6 \cdot  10^{-7}$     \\
\hline 
0.02    & $9.6 \cdot  10^{-8}$     \\
\hline 
0.05    & $4.2 \cdot  10^{-8}$     \\
\hline 
0.1     & $2.0 \cdot 10^{-8}$        \\
\hline
0.15    & $1.6 \cdot 10^{-8}$      \\
\hline
0.2     & $1.2 \cdot 10^{-8}$       \\
\hline
0.25    & $9.8 \cdot 10^{-9}$       \\
\hline
0.5     & $6.0 \cdot 10^{-9}$  \\
\hline
1         & $3.8 \cdot 10^{-9}$        \\
\hline
2        & $4.8\cdot 10^{-9}$         \\
\hline
3        & $1.3 \cdot 10^{-8}$         \\ 
\hline  
5       & $8.4 \cdot 10^{-8}$      \\
\hline
10      & $1.4 \cdot 10^{-6}$       \\
\hline
\end{tabular}
\end{center}
\caption{ Bounds on 
 the value of $\mu_{tran}/\mu_{B}$ for 
different $\nu_{s}$ neutrino masses obtained for case of small mixing angles
 ($\nu_{\mu}\approx \nu_2$, $\nu_s\approx \nu_4$).\ See text.}
\end{table}
    
 \begin{figure}[hbt]
   \mbox{\epsfig{file=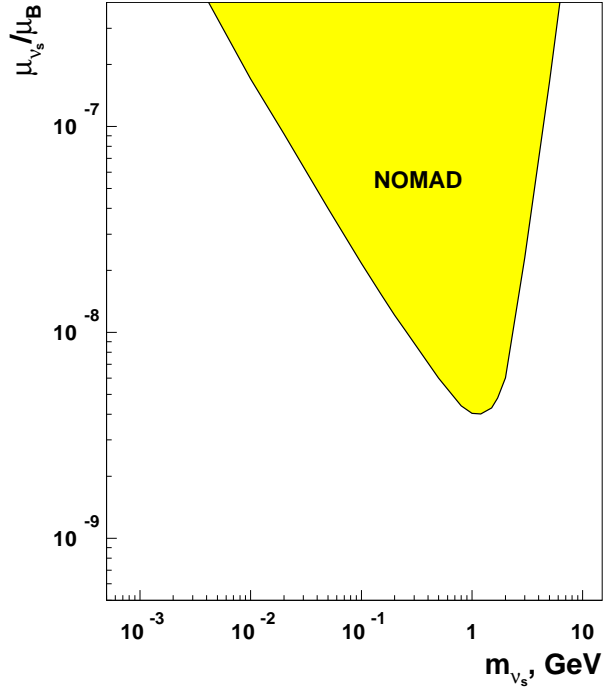,height=100mm}}
    \centering
  \caption{\em The NOMAD exclusion region (dark shaded area)
 in the ($\mu_{\nu_{s}}/\mu_{B}, m_{\nu_{s}}$) parameter space.\
} 
  \label{figure 2:}
\end{figure}

For the two photon process $\nu_{\mu}Z \rightarrow \nu_{\tau}(\nu_{s}) \gamma Z$ 
we have found that the nuclei form-factor gives the main contribution and 
$\int F^2(t)dt \approx d$.\ After numerical calculation we find that 
\begin{equation}
\sigma(\nu_{\mu} Z \rightarrow \nu_{\tau}(\nu_{s})\gamma Z) \approx 
 \frac{0.1~mb}{M^6_{eff}}
\end{equation}

Using the bound of Eq.(18) the parameter 
of the effective Lagrangian, Eqs.(8,13), can be constrained to 
\begin{equation}
\frac{1}{M^6_{eff}} \leq 10^{-11} GeV^{-6},
\end{equation}
for $m_{\nu_{s}} \leq 1GeV$.\ 
From the bounds of Eqs. (18,25) we find that 
\begin{equation}
\Gamma(\nu_{\tau}(\nu_{s}) \rightarrow \nu_{\mu} \gamma \gamma) \leq 
\frac{1}{2\cdot10^{13}sec}m^7_{\nu_{\tau}}(MeV)
\end{equation}

This result in the limit on two-photon neutrino decay time 
\begin{equation}
\tau_{\nu_{\mu}\gamma\gamma} > 2\times10^{13} sec/m_{\nu}^{7}(MeV)
\end{equation}

which  is valid for the $m_{\nu}$ ranges from arbitrary small  masses
up to the value of $m_{\nu}\sim O(1) GeV$, which is defined by $t^2$
suppresion  of the Primakoff cross section (6) at corresponding momentum
transfer.\    
It is interesting to compare this limit with 
 the astrophysical bounds on 
$\nu_{\tau} \rightarrow \nu \gamma$ decay  lifetime for
$m_{\nu_{\tau}} \leq 50~MeV$ \cite{33}:
\begin{equation}
\tau_{\nu\gamma} > 8.4\times10^{8}\Big(MeV/m_{\nu_{\tau}}\Bigr) sec
\end{equation}
or with the BEBC model-independent constraint on limit on
$\nu_{\tau} \rightarrow \nu e^{+} e^{-}$ decay lifetime \cite{21}:
\begin{equation}
\tau_{\nu e^{+} e^{-}} > 0.18 \Big(m_{\nu_{\tau}}/MeV\Bigr) sec
\end{equation}
The limit of Eq.(27) is much stronger.\

For the case of $\nu_{\tau}(\nu_{s}) 
\rightarrow \nu_{e} \gamma \gamma$ decay we 
have a bound that is two orders of magnitude weaker,
 since the fraction of $\nu_{e}$ in the SPS neutrino beam
is $\sim 1\%$.\ Note that 
one can derive analogous bounds on two photon neutrino decay width for 
the case of arbitrary relations between neutrino masses, also  
it is possible to obtain the corresponding bound 
on neutrino-photon cross section from the limit of Eq.(25).

\section {Conclusion}

 It is shown that  non-zero transition magnetic moment 
between 
sterile neutrino  and muon neutrino species could be effectively searched for via the Primakoff effect, 
in the process of $\nu_{\mu} Z \rightarrow \nu_{s}Z$
 conversion in the external Coulomb field of a nucleus, with the subsequent 
 $\nu_{s}\rightarrow \nu_{\mu} + \gamma$ decay.\  From
the recent results of the NOMAD neutrino detector at CERN  
a model-independent constraint of 
\begin{equation}
\mu_{tran} < (10^{-6} - 10^{-9}) \mu_{B}
\end{equation}

is obtained depending on the value of $\nu_{s}$ mass.\  For $m_{s}\sim O(1)~ GeV$ these bounds are comparable with
 the present experimental ones on $\nu_{\mu}$ and 
$\nu_{e}$ diagonal magnetic moments and are a few orders of magnitude more sensitive than those on $\nu_{\tau}$ magnetic moment
 obtained in ref. \cite{20,22,23}.

From the same analysis a constraint on lifetime
 for two-photon neutrino decay $\nu_{\tau}(\nu_{s})\rightarrow \nu_{\mu} +\gamma + \gamma$ 
\begin{equation}
\tau_{\nu_{\mu}\gamma\gamma} > 2\times10^{13} sec/m_{\nu}^{7}(MeV)
\end{equation}
  is obtained.\
The limit is valid for the $m_{\nu}$ range from arbitrary small neutrino masses
up to $m_{\nu}\sim O(1)~GeV$.\
 This limit 
is much more stringent than the bound on radiative $\nu_{\tau} \rightarrow \nu \gamma$ decay  lifetime  
found in ref. \cite{33} or the limit on 
$\nu_{\tau} \rightarrow \nu e^{+} e^{-}$ decay lifetime from ref.\cite{21}.\ 

It should be noted that our estimates based on NOMAD data are rather 
conservative.\ They are obtained for the light target ( NOMAD preshower)
and for the short neutrino decay length.\ However,
one can consider sterile neutrino production in the SPS neutrino beam dump region 
(improvement factor $>10^{4}$ in a probability of $\nu_{\mu}\rightarrow\nu_{s}$ conversion) 
with the subsequent sterile neutrino decay in the full NOMAD fiducial volume
(improvement factor $>500$ in probability of decay).\ Therefore, we believe
the limits o Eqs.(30,31) can be significantly 
improved by more detailed analysis of the NOMAD neutrino data,
 especially for the long-lived (light) sterile neutrinos.

\vspace{1.0cm}

{\large \bf Acknowledgements}\\

We should like to thank V.A. Matveev and V.A. Rubakov for interesting discussions.\ We are indebted to A.V. Kovzelev, M.M. Kirsanov and A.N. Toropin for help in simulation and useful 
comments.\ One of the authors (S.N.G.)
would like to thank his NOMAD colleagues for fruitful discussions,
comments and support.

\end{document}